# Simulation of Phonon-Polariton Generation and Propagation in Ferroelectric LiNbO$_3$ Crystals

David W. Ward, Eric Statz, Nikolay Stoyanov, and Keith A. Nelson
The Massachusetts Institute of Technology
Cambridge, MA 02139, USA

## ABSTRACT

We simulate propagation of phonon-polaritons (admixtures of polar lattice vibrations and electromagnetic waves) in ferroelectric LiNbO$_3$ with a model that consists of a spatially periodic array of harmonic oscillators coupled to THz electromagnetic waves through an electric dipole moment. We show that when this model is combined with the auxiliary differential equation method of finite difference time domain (FDTD) simulations, the salient features of phonon-polaritons may be illustrated. Further, we introduce second order nonlinear coupling to an optical field to demonstrate phonon-polariton generation by impulsive stimulated Raman scattering (ISRS). The phonon-polariton dispersion relation in bulk ferroelectric LiNbO$_3$ is determined from simulation.

## INTRODUCTION

We consider electromagnetic wave propagation (1-dimensional and scalar for simplicity) in crystalline LiNbO$_3$, in which there is an optic phonon resonance of the form,

$$\varepsilon(\omega) = \varepsilon'_\infty + \frac{\omega_{TO}^2 \left( \varepsilon_0 - \varepsilon'_\infty \right)}{\omega_{TO}^2 - \omega^2 - i\omega\Gamma}, \tag{1}$$

where $\omega$ is the electromagnetic wave frequency, $\varepsilon_0$ is the low frequency permittivity, $\varepsilon'_\infty$ is the high frequency permittivity, $\omega_{TO}$ is the transverse optic phonon frequency, and $\Gamma$ is a phenomenological damping rate [1]. Introducing the permittivity in equation (1) into the expression for the macroscopic linear polarization $P(\omega) = \in_0 \left( \varepsilon_r(\omega) - 1 \right) E(\omega)$, we identify an auxiliary parameter $Q$ as follows:

$$P(\omega) = \in_0 \left( \varepsilon'_\infty - 1 \right) E(\omega) + \frac{\in_0 \omega_{TO}^2 \left( \varepsilon'_0 - \varepsilon'_\infty \right)}{\omega_{TO}^2 - \omega^2 - i\omega\Gamma} E(\omega) = \in_0 \left( \varepsilon'_\infty - 1 \right) E(\omega) + \omega_{TO} \sqrt{\in_0 \left( \varepsilon'_0 - \varepsilon'_\infty \right)} Q(\omega). \tag{2}$$

$Q$ may be evaluated independently of equation (2) to determine the total macroscopic polarization by Fourier transformation and solution in the time domain:

$$\frac{d^2 Q}{dt^2}(t) + \Gamma \frac{dQ}{dt}(t) + \omega_{TO}^2 Q(t) = \omega_{TO} \sqrt{\in_0 \left( \varepsilon'_0 - \varepsilon'_\infty \right)} E(t). \tag{3}$$

The coupling constant, appearing as a coefficient to $Q$, is determined by explicitly specifying a model for the mechanical component of the polarization [2]. We consider a transverse optic phonon mode in ferroelectric LiNbO$_3$ scaled by the oscillator density $N$, and the reduced mass of the unit cell $\mu$:

$$Q = \sqrt{N\mu} \left( x^+ - x^- \right), \tag{4}$$

where $x^+ - x^-$ is the ionic displacement of the normal mode. The system can then be thought of as a lattice of radiating harmonic oscillators coupled to each other only through the electromagnetic radiation they generate when oscillating. Introducing equation (3) into Maxwell's equations through the electric displacement field, the complete system of equations (the coupled electro-mechanical system) may be solved numerically with a FDTD simulation and the auxiliary parameter $Q$ [3].

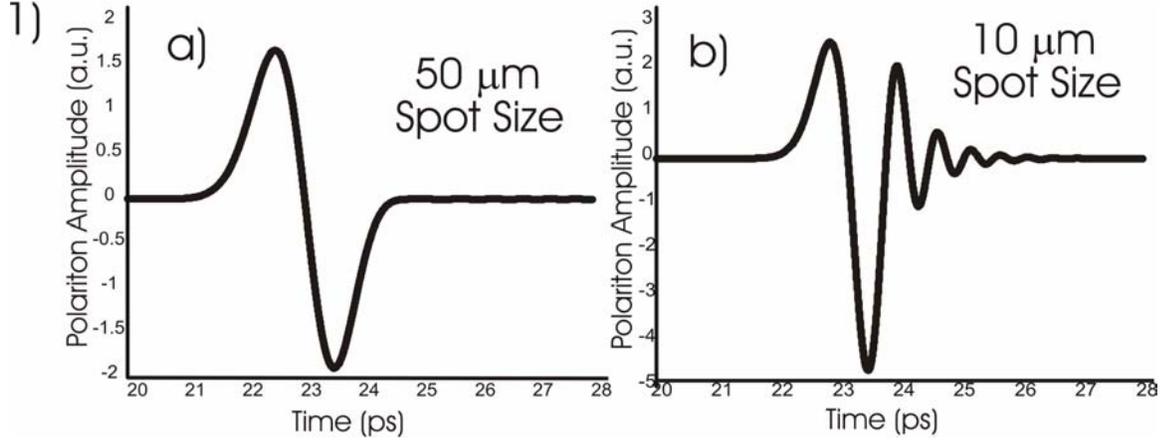

**Figure 1:** Phonon-polaritons generated through ISRS by a 50 fs optical pulse with spot size a) 50 μm and b) 10 μm, and detected after having propagated 1.5 mm away from the excitation region. The polariton wavelength range is dictated by the optical pulse spot size. The polariton response is seen to be virtually non-dispersive in (a), while displaying increasing dispersive properties as the wavelength is decreased.

**SIMULATION**

Solving the complete system of equations using the FDTD method allows for simulation of problem spaces comparable to experiments performed within our research group. These experiments are performed in crystals with volumes ~25 mm$^3$, with phonon-polariton wavelengths on the order of 10-100 μm [4].

Simulations are run on a 24-node Beowulf cluster, each node having ~1.5 GB of dedicated RAM. Communication between the nodes is achieved using 100BT Ethernet interconnects, with gigabit Ethernet connections to a fileserver which manages the simulation output files.

The physical constants of the lowest frequency A1 normal mode in LiNbO3 were taken from Raman and IR studies ($\omega_{TO}$=7.44 THz, $\varepsilon'_0$=4.6, $\varepsilon'_\infty$=20.6, and $\Gamma$=0.844 THz) [5]. The differential polarizability may be calculated from tabulated electro-optic coefficients or measured directly; we use $d\alpha/dx = 8.7x10^{-19}$ m$^2$. The oscillator density is calculated from the unit cell volume: $N = 6.285$ x $10^{27}$ m$^{-3}$.

**RESULTS AND DISCUSSION**

**Polariton Generation**

Polariton generation in LiNbO$_3$ with an ultrashort optical pulse is mediated by impulsive stimulated Raman scattering (ISRS), an off-resonance, nonlinear coupling of the optical radiation to a Raman (and in this case, IR) active phonon mode in the crystal. Momentum is imparted impulsively to this mode through ISRS [6]. The force experienced by the ions due to ISRS is

$$F' = \frac{d^2Q}{dt^2}(t) = \frac{1}{2}\sqrt{\frac{N}{\mu}}\frac{\partial\alpha}{\partial x}|E_{800nm}|^2, \qquad (5)$$

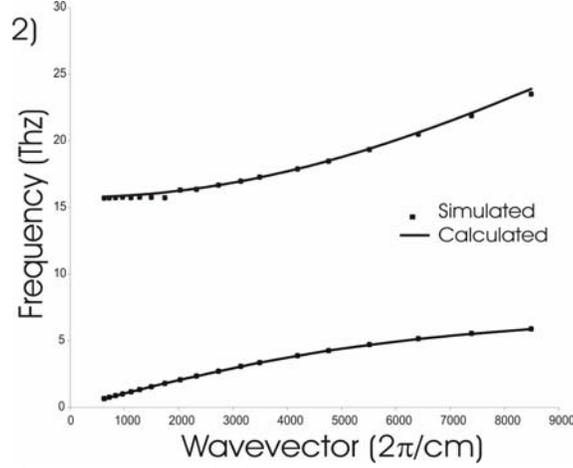

**Figure 2:** Polariton dispersion curve for LiNbO$_3$, generated using 1D FDTD simulations of a 50 fs FWHM excitation pulse.

where $d\alpha/dx$ is the differential polarizability, $E_{800nm}$ is the optical pump field, and the force $F'$ is in scaled units to match those of $Q$. By including this force as a second driving term in equation (3), we can simulate polariton generation.

Figure 1 illustrates the results of polariton generation in two different frequency regimes for broadband pulses. In the linear dispersion regime, which corresponds to frequencies much lower than $\omega_{TO}$ (~0.1-1.0 THz in LiNbO$_3$), a broadband polariton plane wave is generated by focusing of the optical excitation beam to a line of finite width at the crystal. The central frequency of the resulting polariton waveform is proportional to the spot size (i.e. the focused line width). Figure 1a shows the vibrational response to an incident 50 fs optical pulse focused to a beam waist of 50 µm, determined at a distance of 1.5 mm away from the excitation region. The resulting polariton waveform has the profile of the spatial derivative of the Gaussian optical pump profile. The same geometry is employed in figure 1b, but with a spot size of 10 µm. The general character of the waveform is similar to that in 1a, but the effects of polariton group velocity dispersion are apparent. Different frequency components of the polariton wavepacket propagate at different speeds, and by the time the wavepacket has moved 1.5 mm away from the excitation region, it has become significantly distorted.

**Polariton Dispersion**

The effect that the excitation spot size has on polariton propagation is characterized by the polariton dispersion curve which may be obtained by introducing the permittivity in equation (1) into the dispersion relation obtained by considering plane wave solutions to Maxwell's equations, illustrated below in the simplified case of zero damping:

$$\frac{c_0^2 k_0^2}{\omega^2} = \varepsilon(\infty) + \frac{\omega_{TO}^2 \left(\varepsilon_0 - \varepsilon_\infty'\right)}{\omega_{TO}^2 - \omega^2}$$

$$\omega(k) = \sqrt{\frac{1}{2}\left(\frac{c_0^2 k^2}{\varepsilon_\infty} + \omega_{TO}^2\left(1 + \left(\varepsilon_0 - \varepsilon_\infty'\right)\right)\right) \pm \frac{1}{2}\sqrt{\left(\frac{c_0^2 k^2}{\varepsilon_\infty} + \omega_{TO}^2\left(1 + \left(\varepsilon_0 - \varepsilon_\infty'\right)\right)\right)^2 + \frac{4c_0^2 k^2}{\varepsilon_\infty}\omega_{TO}^2\left(\varepsilon_0 - \varepsilon_\infty'\right)}}, \tag{6}$$

where $c_0$ is the speed of light in vacuum. There are two sets of solutions, the lower and upper polariton branches. The solution of equation (6) is illustrated in figure 2, where it is compared to simulation. The simulated result includes a phenomenological damping rate, but this does not alter the resonance frequencies substantially as evidenced in the figure.

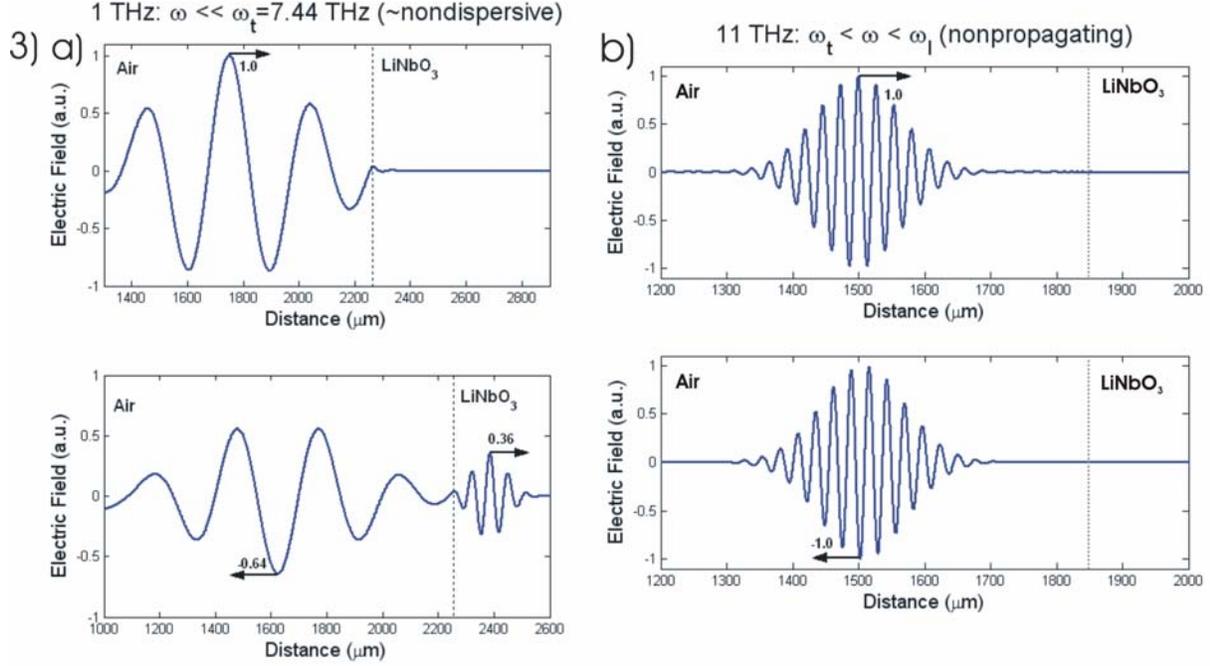

**Figure 3:** THz radiation enters LiNbO₃ from the left and reflects and transmits at the interface (reflection and transmission coefficients are indicated in the figure). a) 1 THz radiation partially transmitting and reflecting within the linear dispersion region of the crystal. b) Total reflection at 11 THz, which is in the bandgap of LiNbO₃.

The spatial intensity distribution of the optical pulse at the sample dictates the wavevector range of the polariton wavepacket that is generated. Alternatively, by using a spatially periodic excitation pattern, specified polariton wavevectors may be selected. Once again, we monitor a fixed point on the spatial grid of the simulation with a well defined separation from the excitation region so that the temporal evolution of the polaritons may be recorded as they pass by.

### Polariton Bandgap

In addition to generating polaritons directly in the host crystal, we also directed THz radiation from free space into a LiNbO₃ crystal. Figure 3a shows a 1 THz, few-cycle polariton pulse after having been reflected from the crystal interface. Transmission is also evident and the reflection and transmission coefficients are indicated in the figure. Figure 3b directs 11 THz radiation at the same interface, but this frequency is in the bandgap of LiNbO₃, i.e. the region between the transverse optical phonon frequency (7.44 THz) and the longitudinal optic phonon frequency given by the Lyddane-Sachs-Teller relation (15.74 THz),

$$\omega_{LO} = \sqrt{\frac{\varepsilon_0'}{\varepsilon_\infty'}}\omega_{TO}. \tag{7}$$

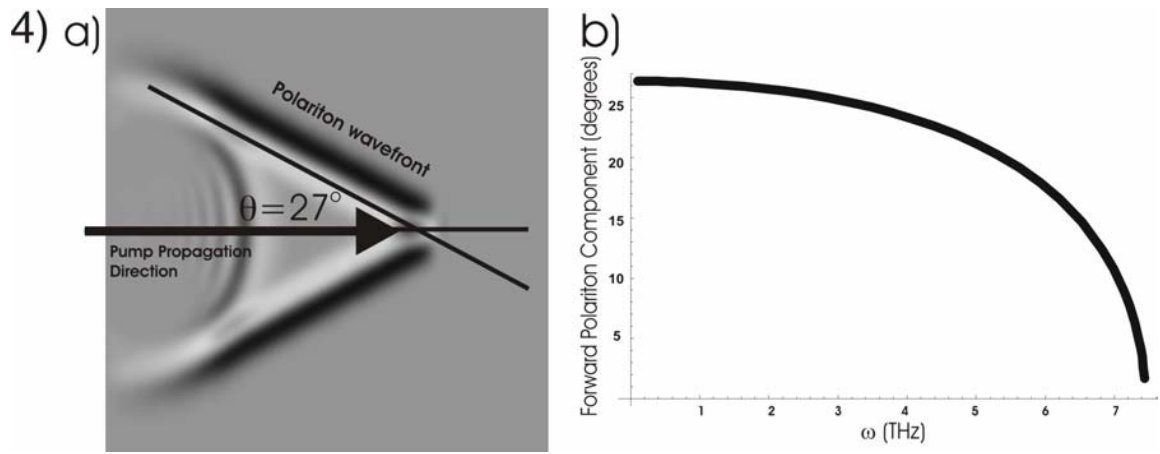

**Figure 4:** a. 2D FDTD simulation of low frequency polariton generation in LiNbO₃, yielding a wavefront angle of 27°. b. Calculated wavefront angle as a function of polariton frequency, in the low frequency limit, 27°.

Close inspection of 11 THz penetration at the interface indicates minimal penetration analogous to skin depth in a metal. Simulations were performed at THz frequencies above the bandgap as well and show a return to propagation.

## Forward Polariton Component

Two-dimensional simulations reveal that as the optical pump pulse propagates through the crystal, the polaritons propagate laterally away from the excitation region; the angle which the polariton wavefront makes with respect to the pump propagation direction is determined by the relative indices of refraction in LiNbO₃ for the optical pump and THz radiation [7]. Figure 4a shows a FDTD simulation result that exhibits the forward component with the angle borne out by theory and experiment. Figure 4b shows the angle calculated from the index of refraction for an 800 nm optical pulse beam ($n$=2.18) and THz permittivity calculated from equation (1) as a function of THz frequency.

## CONCLUSION

We have employed FDTD simulations with an auxiliary parameter integrated independently to model phonon-polaritons as harmonic oscillators coupled to an electromagnetic wave. The simulation parameters were set to identify the mechanical component with a transverse optic phonon normal mode in LiNbO₃ that has been explored experimentally by our research group. We have shown that the model is consistent with the properties of ferroelectric LiNbO₃. It is not necessary to assume only a harmonic oscillator. The auxiliary parameter may be expanded to include higher order terms to allow for simulation of nonlinear lattice vibrations and their effect on polariton propagation. Since the numerical solutions were pursued with FDTD, it is also not necessary to restrict problems to isotropic materials, but the anisotropic features of ferroelectrics may also be simulated. Similarly, inhomogeneous materials, metals, and semiconductor polariton interactions may also be simulated with this method.

**ACKNOWLEDGEMENTS**

This work was supported in part by the Cambridge-MIT Institute Grant No. CMI-001.